\input amstex
\documentstyle{amsppt}

\topmatter
\title A Rigorous Real Time Feynman Path Integral
       \endtitle
\author Ken Loo \endauthor
\address P.O. Box 9160, Portland, Oregon 97207\endaddress
\email look\@sdf.lonestar.org\endemail
\dedicatory This work partially fulfills the author's
   Ph.D. thesis requirements
   under the guidance of Professor James Thurber at Purdue
   University.  This work was supported by the Purdue
   research foundation.
   \enddedicatory
\thanks The author would like to give thanks to his Ph.D. thesis committee:
        Daniel Gottlieb, Leonard Lipshitz, Herman Rubin, and
        James Thurber.  Also, special thanks to Patricia Bauman. 
        \endthanks
\abstract Using improper Riemann integrals, 
   we will formulate a rigorous version of the
   real-time, time-sliced Feynman path integral for 
   the $L^2$ transition probability amplitude.  We will
   do this for nonvector potential Hamiltonians with 
   potential which has at most a finite number of 
   discontinuities and singularities.  We will also provide
   a Nonstandard Analysis version of our formulation.  
\endabstract
\endtopmatter
\document

\define\fac#1#2{\left(\frac {m}{2\pi{}i\hbar\epsilon}\right)^
    {\frac{#1}2\left(#2 + 1 \right)}}

\define\summ#1#2#3#4{\frac {i\epsilon}{\hbar}\sum\limits_{j=1}
    ^{{#1}+1}\left[\frac {m}{2}\left(\frac {{#2} - {#3}}
    {\epsilon}\right)^2\! #4\right]}

\define\pd#1#2{\dfrac{\partial#1}{\partial#2}}
\define\spd#1#2{\tfrac{\partial#1}{\partial#2}}
\define\ham1#1#2{\dfrac{-\hbar^2}{2m}\Delta_{#1} #2}

\subhead
{\bf 1. Introduction and Notations}\endsubhead
In this paper, we will formulate a rigorous version of the
real-time, time-sliced Feynman path integral for the
$L^2$ transition probability amplitude
$$
  \Bigg<\phi^{*},\,
       \text{exp}\left(\dfrac{-it\bar H}{\hbar}\right)
       \psi\Bigg>_{L^2} = 
  \int_{\Bbb R^n} \phi\left(\vec x\right)\left[
  \text{exp}
  \left(\dfrac{-it\bar H}{\hbar}\right)\psi\right]\left(\vec x\right)
  d\vec x , \tag1.1
$$
where $\phi ,\psi\in L^2$, 
$H = \dfrac{-\hbar^2}{2m}\Delta + V\left(\vec x\right)$ is 
essentially self-adjoint, $\bar H$ is the closure
of $H$, and $\phi ,\psi , V$ each carries at most a finite number
of singularities and discontinuities.  In flavor of physics
literature, we will formulate the Feynman path integral with
improper Riemann integrals.  In hope that with further research
we can formulate a rigorous 
polygonal path integral, we will also provide
a Nonstandard Analysis version of the Feynman path integral.  
Using Nonstandard Analysis is not essential to our formulation,
and the idea of using Nonstandard Analysis on the Feynman path
integral is not a new concept. 
For readers interested in Nonstandard Analysis, and its 
applications to Feynman
path integrals, see [1], [10], [13], [19], 
[22], and references within.  We will
assume that the reader is familiar with Nonstandard Analysis.  

In physics, the Feynman path integral is formulated on the
propagator and it is formally given by
(see [11], [14], and [21]) 
$$
  K_t\left(\vec x, \vec x_0\right) = \lim_{k\to\infty}
  w_{n,k}\int_{\Bbb R^{\left(k-1\right)n}}
  \text{exp}\left[\dfrac{i\epsilon}{\hbar}
   S_k\left(\vec x,...\vec x_o\right)\right]
   d\vec x_1...d\vec x_{k-1},\tag1.2$$
where
$$\align
   {}&w_{n,k} = \left(\dfrac{m}{2i\pi\hbar\epsilon}\right)^
    \frac{nk}{2}, \epsilon = \dfrac{t}{k}, \tag1.3\\
   {}&S_k\left(\vec x = \vec x_k,\dots ,\vec x_0\right) =
   \sum\limits_{j=1}
   ^{k}\left[\dfrac {m}{2}\left(\dfrac {\vec x_j - \vec x_{j-1}}
   {\epsilon}\right)^2\! - V\left(\vec x_j\right)\right].
\endalign
$$
and all integrals are improper Riemann integrals.

In mathematics, there is a there is a rigorous time-sliced
Feynman path integral for the wave function
(see [5] and [18])
$$\align
{}&\left[
   \text{exp}\left(\dfrac{-it\bar H}{\hbar}\right)
   \psi\right]\left(\vec x\right) = \tag1.4 \\
   {}&\text{l.i.m}_{k\to\infty}w_{n,k}
   \int_{\Bbb R^{kn}}
   \text{exp}\left[\dfrac{i\epsilon}{\hbar}
   S_k\left(\vec x,\dots,\vec x_o\right)
   \right]\psi\left(\vec x_0\right)
   \,d\vec x_0\dots d\vec x_{k-1}.
\endalign
$$
where the integrals in (1.4) are improper
Lebesgue integrals and their convergence
is in the $L^2$ norm.  
Other popular rigorous versions of the Feynman path
integral are the Wiener integral(see [7], [8], [12],
[16], and [18]),
generalization of Fresnel integrals(see [2]),
and Henstock integrals(see [15]).
For a more detailed exposition and further references,
see [2] and [3].  

Our main concern in this paper is to provide a rigorous
version of (1.2) for the transition probability
amplitude given in (1.1) by using (1.4).  We will 
show that for any essentially self-adjoint Hamiltonian
with potential that carries a finite number of singularities
and discontinuities and for any $\phi ,\psi\in L^2$ which
also has a finite number of singularities and discontinuities
the following holds 
$$\align
  {}&\int_{\Bbb R^n} \phi\left(\vec x\right)\left[
  \text{exp}
  \left(\dfrac{-it\bar H}{\hbar}\right)\psi\right]\left(\vec x\right)
  d\vec x  = \tag1.5 \\
  {}&\lim_{k\to\infty}
    w_{n,k}
   \int_{\Bbb R^{\left(k+1\right)n}}\phi\left(\vec x_k\right)
   \text{exp}\left[\dfrac{i\epsilon}{\hbar}
   S_k\left(\vec x_k,\dots ,\vec x_0\right)
   \right]\psi\left(\vec x_0\right)
   \,d\vec x_0\dots d\vec x_{k}. 
\endalign
$$
In the last line of (1.5), 
the integral is an improper Riemann integral over
$\Bbb R^{\left(k+1\right)n}$.  

A trivial application of Nonstandard Analysis on the $k$
limit in (1.5) yields
$$\align
  {}&\int_{\Bbb R^n} \phi\left(\vec x\right)\left[
  \text{exp}
  \left(\dfrac{-it\bar H}{\hbar}\right)\psi\right]\left(\vec x\right)
  d\vec x  = \tag1.6 \\
  {}&st\Bigg\{w_{n,\omega}
   \int_{{}^{{}^*}
   \Bbb R^{\left(\omega +1\right)n}}\phi\left(\vec x_\omega\right)
   \text{exp}\left[\dfrac{i\epsilon}{\hbar}
   S_\omega\left(\vec x_\omega ,\dots ,\vec x_0\right)
   \right]\psi\left(\vec x_0\right)
   \,d\vec x_0\dots d\vec x_{\omega}\Bigg\},
\endalign
$$
where the integral in the last
line of (1.6) is a $*$-transformed improper Riemann
integral over ${}^{^{*}}\Bbb R^{\left(\omega +1\right)n}$,
and $\omega\in {}^*\Bbb N - \Bbb N$.  

The main idea in the proof of (1.5) 
is the following.  For simplicity, 
suppose $f\left(x\right)\in L^2\left(\Bbb R\right), 
g\left(x, y\right)
\in L^2\left(\Bbb R\times\Bbb R\right)$ are such that
they are bounded and continuous.  Further, suppose that
both 
$$\align
  {}&h\left(x\right) = \int_{-a}^{b}
   g\left(x, y\right)dy,  \tag1.7\\
  {}&p\left(x\right) = 
   \text{l.i.m}_{a,b\to\infty}\int_{-a}^{b}
   g\left(x, y\right)dy
\endalign
$$
are in $L^2\left(\Bbb R\right)$ as a function of
$x$.  In (1.7), we take the integral to
be Lebesgue integrals and the limits are taken
independent of each other.  
Notice that for $p\left(x\right)$, we can interpret the
integral 
as an improper Lebesgue integral with convergence
in the $L^2$ topology.  Let us denote $\chi_{[-c,d]}$
to be the characteristic function on $[-c, d]$. 
Schwarz's inequality then implies 
$$\align
  {}&\Bigg|\int_{\Bbb R} f\left(x\right)p\left(x\right) -
  \int_{-c}^{d}
  \int_{-a}^{b}
  f\left(x\right)g\left(x, y\right)dx dy\Bigg| \leq \tag1.8\\
  {}&||f||_2\,||p - h||_2 + ||f - \chi_{[-c,d]}f||_2\,||h||_2
    \to 0. 
\endalign
$$
\comment
  \int_{\Bbb R}\left\{\text{l.i.m}_{c,d\to\infty}
  \chi_{[-c,d]}f\left(x\right)\right\}
  h\left(x\right) = \\
  {}&\lim_{a,b,c,d\to\infty}\int_{-c}^{d}
  \int_{-a}^{b}f\left(x\right)g\left(x, y\right)dx dy
\endcomment
Thus, we can write
$$
  \int_{\Bbb R} f\left(x\right)p\left(x\right) =
  \lim_{a,b,c,d\to\infty}\int_{-c}^{d}
  \int_{-a}^{b}f\left(x\right)g\left(x, y\right)dx dy, \tag1.9
$$
where the limits are all taken independent of each other.
Since $f$ and $g$ are bounded and continuous,
the Lebesgue integral over $[-a, b]\times [-c, d]$
in (1.9) can be replaced by
a Riemann integral.  Since the limits are taken independent
of each other, we can then interpret the right hand-side 
of (1.9) as an improper Riemann integral.  If $f$ and
$g$ carry singularities and discontinuities, 
care must be taken in the
region of integration so that
the replacement of Lebesgue integral with Riemann integrals
can be done.

We now set some notations to deal with $n$-dimensional 
integrations, singularities and discontinuities.    
Let $k\in \Bbb N$ and $0\leq l\leq k$.  We will
denote the interior of the $l$th box by  
$$ 
  A^l = (-a_1^l, b_1^l)\times\dots\times (-a_n^l, b_n^l),\tag1.10
$$
for positive and large $a$'s and $b$'s.  
Let $K = \left\{\vec y_1\dots\vec y_p\right\}$
be the set of discontinuous and singular
points of $\phi ,\psi$ and $V$.  For each
$\vec y_q = (y_1^q,\dots ,y_n^q)\in K$, denote
the $l$th box centered at $\vec y_q$ by 
$$
  B_q^l = (y_1^q - \frac{1}{c_1^{q,l}},\,  
   y_1^q + \frac{1}{d_1^{q,l}})\times\dots\times
   (y_n^q - \frac{1}{c_n^{q,l}},\,  
   y_n^q + \frac{1}{d_n^{q,l}}), \tag1.11
$$
for positive and large $c$'s and $d$'s.
Let 
$$
  C^l = A^l - \left\{\bigcup_{q = 1}^{p}B_q^l\right\}.\tag1.12
$$
For arbitrary large $a$'s, $b$'s, 
$c$'s, and $d$'s, $C^l$ is a box which encloses
the set $K$ and at each point of $K$, a small
box centered at that point is taken out.  
Associated with $C^l$ is a set of indices 
$$\align
   \left\{j_l\right\} = \{a_1^l,\dots ,a_n^l, b_1^l,\dots ,b_n^l,
    {}&c_1^{1,l},\dots ,c_n^{1,l}, \dots ,
    c_1^{p,l},\dots ,c_n^{p,l}, \tag1.13\\
    {}&d_1^{1,l},\dots ,d_n^{1,l}, \dots ,
    d_1^{p,l},\dots ,d_n^{p,l}\}
\endalign
$$
We will denote by $\left\{j_l\right\}\to\infty$ to
mean
$$\align
    {}&a_1^l,\dots ,a_n^l, b_1^l,\dots ,b_n^l,
    c_1^{1,l},\dots ,c_n^{1,l}, \dots ,
    c_1^{p,l},\dots ,c_n^{p,l},\tag1.14\\
    {}&d_1^{1,l},\dots ,d_n^{1,l}, \dots ,
    d_1^{p,l},\dots ,d_n^{p,l}\to\infty ,
\endalign
$$
where all indices goes to infinity independent
of each other.  Notice that as 
$\left\{j_l\right\}\to\infty$,
we recover $\Bbb R^n$ $a.e.$ from $C^l$.  
We will denote by $\chi_{\left\{j_l\right\}}$ the characteristic
function on $C^l$.  Notice that 
for $f\in L^2\left(\Bbb R^n\right)$,
$$
 \text{l.i.m}_{\left\{j_l\right\}\to\infty}
 \chi_{\left\{j_l\right\}}f = f \quad a.e. \tag1.15
$$

Lastly, let us write
$$
  D_{\left\{J_0^k\right\}} = 
  C^0\times\dots\times C^k. \tag1.16
$$
Associated with $D_{\left\{J_0^k\right\}}$
is a set of indices
$$
  \left\{J_0^k\right\} = 
  \bigcup_{l = 0}^{k}\left\{j_l\right\}, \tag1.17
$$
and as before, we will use the notation
$\left\{J_0^k\right\}\to\infty$ to mean
$$
  \left\{j_0\right\}\to\infty ,\dots 
  ,\left\{j_k\right\}\to\infty , \tag1.18
$$
where the indices are taken to infinity
independent of each other.

From here on, we will assume that
$\phi ,\psi \in L^2$ and $V$ are such
that they have at most a finite
number of singularities and discontinuities
and the set of those points are denoted as
$K = \left\{\vec y_1\dots\vec y_p\right\}$. 
Finally, we will denote by $\int_{rO}$ to be
Riemann or improper Riemann integration over
the region $O$ and $\int_{O}$ to be
Lebesgue integration over the region $O$.

\subhead
{\bf 2. Feynman Path Integrals}
\endsubhead
The standard derivation of (1.4) is via
the Trotter product formula
(see [5], [16], and [18]) which
says that for any essentially self-adjoint
$H = H_0 + V$ with 
$H_0 = \dfrac{-\hbar^2}{2m}\Delta$ and
any $\psi\in L^2$,
$$
  \text{exp}\left(\dfrac{-it\bar H}{\hbar}\right)\psi =
  \text{l.i.m.}_{k\to\infty}\left\{
  \text{exp}\left(\dfrac{-itV}{k\hbar}\right)
  \text{exp}\left(\dfrac{-itH_0}{k\hbar}\right)\right\}^k\psi. \tag2.1
$$
Thus, we have the following
\proclaim{\bf Lemma 2.1}Suppose 
$H = H_0 + V$ is essentially self-adjoint.
Let $\psi , \phi\in L^2$, then
$$\align
  {}&\int_{\Bbb R^n} \phi\left(\vec x\right)
  \left[\text{exp}\left(\dfrac{-it\bar H}{\hbar}\right)
  \psi\right]\left(\vec x\right)\,d\vec x = \tag2.2\\
  {}&\lim_{k\to\infty}
  \int_{\Bbb R^n} \phi\left(\vec x\right)
  \left[\left\{\text{exp}\left(\dfrac{-itV}{k\hbar}\right)
  \text{exp}\left(\dfrac{-itH_0}{k\hbar}\right)
  \right\}^k\psi
  \right]\left(\vec x\right)\,d\vec x\,.
  \endalign
$$
\endproclaim
\demo{Proof} The proof is just an application 
of (2.1) and Schwarz's inequality. \qed
\enddemo

It is well known that(see [5] and [18]) 
for $\nu\in L^1\left(\Bbb R^n\right)\bigcap
L^2\left(\Bbb R^n\right)$,
$$
  \left[\text{exp}\left(\dfrac{-i\epsilon{} H_0}{\hbar}\right)
  \nu\right]\left(\vec x_1\right) =
  \left(\dfrac{m}{2i\pi\hbar\epsilon}\right)^
  \frac{n}{2}\int_{\Bbb R^n} \text{ exp }\left[\dfrac{im\epsilon}{2\hbar}
  \left(\dfrac{\vec x_1 - \vec x_0}{\epsilon}\right)^2
  \right]
  \nu\left(\vec x_0 \right)\, d\vec x_0 , \tag2.3
$$
and that the operator 
$\text{exp}\left(\dfrac{-i\epsilon{} H_0}{\hbar}\right)$
is unitary.  Thus, we can write
$$\align
  {}&\left[\text{exp}\left(\dfrac{-i\epsilon{} H_0}{\hbar}\right)
  \psi\right]\left(\vec x_1\right) =
  \left[\text{exp}\left(\dfrac{-i\epsilon{} H_0}{\hbar}\right)
  \left[\text{l.i.m}_{\left\{j_0\right\}\to\infty}
  \chi_{\left\{j_0\right\}}\psi
  \right]
  \right]\left(\vec x_1\right) = \tag2.4\\
  {}&\text{l.i.m}_{\left\{j_0\right\}\to\infty}
     \left[\text{exp}\left(\dfrac{-i\epsilon{} H_0}{\hbar}\right)
     \chi_{\left\{j_0\right\}}\psi\right]\left(\vec x_1\right) = \\ 
  {}&\text{l.i.m}_{\left\{j_0\right\}\to\infty}
     \left(\dfrac{m}{2i\pi\hbar\epsilon}\right)^
  \frac{n}{2}\int_{C^0} \text{ exp }\left[\dfrac{im\epsilon}{2\hbar}
  \left(\dfrac{\vec x_1 - \vec x_0}{\epsilon}\right)^2
  \right]
  \psi\left(\vec x_0 \right)\, d\vec x_0.
\endalign
$$
Notice that by construction of the region
$C^0$, $\psi$ is bounded and continuous
on $C^0$, hence the Lebesgue integral   
in the last line of (2.4) can be replaced
by a Riemann integral.  

For notation convenience, we will denote  
$$\align
  {}&\rho\left(\vec x_{k}, 
     \left\{J_0^{k-1}\right\}\right) = \tag2.5\\
  {}&w_{n,k}
     \int_{D_{\left\{J_0^{k-1}\right\}}}
     \text{exp}\left[\dfrac{it}{\left(k+1\right)\hbar}
     S_k\left(\vec x_k, \dots ,\vec x_o\right)
     \right]\psi\left(\vec x_0\right)
     \,d\vec x_0\dots d\vec x_{k-1}, \\  
  {}&T = \text{exp}\left(\dfrac{-itV}{k\hbar}\right)
  \text{exp}\left(\dfrac{-itH_0}{k\hbar}\right), \quad 
  T^k = \left\{\text{exp}\left(\dfrac{-itV}{k\hbar}\right)
  \text{exp}\left(\dfrac{-itH_0}{k\hbar}\right)
  \right\}^k.
\endalign
$$
\proclaim{\bf Lemma 2.2}Suppose
$H = H_0 + V$ is essentially self-adjoint.
Let $\psi\in L^2$, then for $k\in\Bbb N$
the following holds
\endproclaim
$$\align
  {}&\left\{\text{exp}\left(\dfrac{-itV}{k\hbar}\right)
  \text{exp}\left(\dfrac{-itH_0}{k\hbar}\right)
  \right\}^k\psi = \tag2.6 \\
  {}&\text{l.i.m}_{\left\{J_0^{k-1}\right\}\to\infty}
     w_{n,k}
     \int_{D_{\left\{J_0^{k-1}\right\}}}
     \text{exp}\left[\dfrac{i\epsilon}{\hbar}
     S_k\left(\vec x_k, \dots ,\vec x_o\right)
     \right]\psi\left(\vec x_0\right)
     \,d\vec x_0\dots d\vec x_{k-1}.
\endalign
$$     
\demo{Proof} We will proof (2.6) by induction.  Suppose
$k = 2$, then (2.4) implies 
$$\align
  {}&T^2\psi = 
  T\left\{\text{l.i.m}_{\left\{J_0^{0}\right\}\to\infty}
  \rho\left(\vec x_{1},
     \left\{J_0^{0}\right\}\right)\right\} = \tag2.7\\
  {}&\text{exp}\left(\dfrac{-itV}{2\hbar}\right)
  \text{exp}\left(\dfrac{-itH_0}{2\hbar}\right)
  \Bigg\{\text{l.i.m}_{\left\{j_1\right\}\to\infty}
  \chi_{\left\{j_1\right\}}\left[
  \text{l.i.m}_{\left\{J_0^{0}\right\}\to\infty}
  \rho\left(\vec x_{1},
     \left\{J_0^{0}\right\}\right)\right]\Bigg\}.
\endalign
$$
Since multiplication by a characteristic
function, 
$\text{exp}\left(\dfrac{-itV}{2\hbar}\right)$,
and $\text{exp}\left(\dfrac{-itH_0}{2\hbar}\right)$
are all continuous operators from $L^2$ to
$L^2$, we can take the $L^2$ limits in (2.7)
outside of the operators and we can do this
in any order we wish. 
Hence, (2.6) is true
for $k = 2$.  Assuming (2.6) to be true for k,
then 
$$\align
  {}&T^{k+1}\psi = 
   \text{exp}\left(\dfrac{-itV}
   {\left(k+1\right)\hbar}\right)\times \tag2.8 \\
  {}&\text{exp}\left(\dfrac{-itH_0}{\left(k+1\right)\hbar}\right)
  \Bigg\{\text{l.i.m}_{\left\{j_{k}\right\}\to\infty}
  \chi_{\left\{j_{k}\right\}}\left[
  \text{l.i.m}_{\left\{J_0^{k - 1}\right\}\to\infty}
  \rho\left(\vec x_{k},
     \left\{J_0^{k-1}\right\}\right)\right]\Bigg\}.
\endalign
$$
By the same reasoning as for the case of $k = 2$, 
we can take all the $L^2$ limits in (2.8) outside
of the operators and we can do this in any order
we wish.  Hence, (2.6) is true for all $k\in\Bbb N$. \qed
\enddemo

\proclaim{\bf Proposition 2.3} Suppose
  $H = H_0 + V$ is essentially self-adjoint.
  Let $\psi , \phi\in L^2$, then for all
  $k\in\Bbb N$ the following is true
$$\align
  {}&\int_{\Bbb R^n} \phi\left(\vec x\right)
  \left[\left\{\text{exp}\left(\dfrac{-itV}{k\hbar}\right)
  \text{exp}\left(\dfrac{-itH_0}{k\hbar}\right)
  \right\}^k\psi
  \right]\left(\vec x\right)\,d\vec x\, = \tag2.9\\
  {}&w_{n,k}
   \int_{r\Bbb R^{\left(k+1\right)n}}\phi\left(\vec x_k\right)
   \text{exp}\left[\dfrac{i\epsilon}{\hbar}
   S_k\left(\vec x_k,\dots ,\vec x_0\right)
   \right]\psi\left(\vec x_0\right)
   \,d\vec x_0\dots d\vec x_{k}. 
\endalign
$$
\endproclaim
\demo{Proof} Lemma 2.2 implies that
$$\align
  {}&\int_{\Bbb R^n} \phi\left(\vec x\right)
  \left[\left\{\text{exp}\left(\dfrac{-itV}{k\hbar}\right)
  \text{exp}\left(\dfrac{-itH_0}{k\hbar}\right)
  \right\}^k\psi
  \right]\left(\vec x\right)\,d\vec x\, = \tag2.10 \\
  {}&w_{n,k}\int_{\Bbb R^n}\Bigg\{ 
     \text{l.i.m}_{\left\{j_k\right\}\to\infty}
     \chi_{\left\{j_k\right\}}
     \phi\left(\vec x_k\right)\Bigg\}\Bigg\{
     \text{l.i.m}_{\left\{J_0^{k-1}\right\}\to\infty}
     w_{n,k} \\
  {}&\int_{D_{\left\{J_0^{k-1}\right\}}}
     \text{exp}\left[\dfrac{i\epsilon}{\hbar}
     S_k\left(\vec x_k, \dots ,\vec x_o\right)
     \right]\psi\left(\vec x_0\right)
     \,d\vec x_0\dots d\vec x_{k-1}\Bigg\}d\vec x_k .
\endalign
$$
We now apply the idea in (1.8) and (1.9). 
Since all limits in (2.10) are taken independent
of each other, we can use Schwarz's inequality
and take all the $L^2$ limits outside of the integral
as pointwise limits.  Thus,
$$\align
  {}&\int_{\Bbb R^n} \phi\left(\vec x\right)
  \left[\left\{\text{exp}\left(\dfrac{-itV}{k\hbar}\right)
  \text{exp}\left(\dfrac{-itH_0}{k\hbar}\right)
  \right\}^k\psi
  \right]\left(\vec x\right)\,d\vec x\, = \tag2.11 \\
  {}&w_{n,k}\lim_{\left\{J_0^k\right\}\to\infty}
     \int_{D_{\left\{J_0^k\right\}}}
     \phi\left(\vec x_k\right)
     \text{exp}\left[\dfrac{i\epsilon}{\hbar}
     S_k\left(\vec x_k, \dots ,\vec x_o\right)
     \right]
     \psi\left(\vec x_0\right)d\vec x_0\dots d\vec x_{k}.\tag2.12
\endalign
$$     
By construction of $D_{\left\{J_0^k\right\}}$,
the integrand $\phi\left(\vec x_k\right)
     \text{exp}\left[\dfrac{i\epsilon}{\hbar}
     S_k\left(\vec x_k, \dots ,\vec x_o\right)
     \right]
     \psi\left(\vec x_0\right)$ in (2.12) is
a bounded and continuous function on 
$D_{\left\{J_0^k\right\}}$.  Hence, we can
replace the Lebesgue integrals in 
(2.12)
by Riemann integrals.  Since all limits
in (2.12) are taken
independent of each other, we can 
interpret (2.12) as an improper Riemann integral. \qed
\enddemo

We are now ready to proof (1.5).
\proclaim{\bf Theorem 2.4} Suppose
  $H = H_0 + V$ is essentially self-adjoint.
  Let $\psi , \phi\in L^2$.  Furthermore,
  suppose that $\psi ,\phi ,$ and $V$
  has at most a finite number of singularities
  and discontinuities.  With our previously defined
  notations, the following is true
$$\align
  {}&\int_{\Bbb R^n} \phi\left(\vec x\right)\left[
  \text{exp}
  \left(\dfrac{-it\bar H}{\hbar}\right)\psi\right]\left(\vec x\right)
  d\vec x  = \tag2.13 \\
  {}&\lim_{k\to\infty}
    w_{n,k}
   \int_{r\Bbb R^{\left(k+1\right)n}}\phi\left(\vec x_k\right)
   \text{exp}\left[\dfrac{i\epsilon}{\hbar}
   S_k\left(\vec x_k,\dots ,\vec x_0\right)
   \right]\psi\left(\vec x_0\right)
   \,d\vec x_0\dots d\vec x_{k}.
\endalign
$$
\endproclaim
\demo{Proof} Follows from lemma 2.1 and
proposition 2.3. \qed
\enddemo

\subhead
{\bf 3. Nonstandard Feynman Path Integrals}\endsubhead
A trivial application of Nonstandard Analysis on the 
$K$ limit in (2.2) will produce (1.6).  It is our
hope that with further research, a rigorous nonstandard
polygonal path integral can be formulated.  

\proclaim{\bf Theorem 3.1} 
  Suppose
  $H = H_0 + V$ is essentially self-adjoint.
  Let $\psi , \phi\in L^2$.  Furthermore,
  suppose that $\psi ,\phi ,$ and $V$
  has at most a finite number of singularities
  and discontinuities.  With our previously defined
  notations, the following is true
$$\align
  {}&\int_{\Bbb R^n} \phi\left(\vec x\right)\left[
  \text{exp}
  \left(\dfrac{-it\bar H}{\hbar}\right)\psi\right]\left(\vec x\right)
  d\vec x  = \tag3.1 \\
  {}&st\Bigg\{w_{n,\omega}
   \int_{r{}^{{}^*}
   \Bbb R^{\left(\omega +1\right)n}}\phi\left(\vec x_\omega\right)
   \text{exp}\left[\dfrac{i\epsilon}{\hbar}
   S_\omega\left(\vec x_\omega ,\dots ,\vec x_0\right)
   \right]\psi\left(\vec x_0\right)
   \,d\vec x_0\dots d\vec x_{\omega}\Bigg\},
\endalign
$$
where the integral in the last
line of (3.1) is a $*$-transformed improper Riemann
integral over ${}^{^{*}}\Bbb R^{\left(\omega +1\right)n}$,
and $\omega\in {}^*\Bbb N - \Bbb N$.
\endproclaim

\demo{Proof} The nonstandard equivalent of lemma 2.1 is that
for any $\omega\in {}^*\Bbb N - \Bbb N$,
$$\align
  {}&\int_{\Bbb R^n} \phi\left(\vec x\right)
  \left[\text{exp}\left(\dfrac{-it\bar H}{\hbar}\right)
  \psi\right]\left(\vec x\right)\,d\vec x = \tag3.2\\
  {}&st\Bigg\{\int_{\Bbb R^n} \phi\left(\vec x\right)
  \left[\left\{\text{exp}\left(\dfrac{-itV}{\omega\hbar}\right)
  \text{exp}\left(\dfrac{-itH_0}{\omega\hbar}\right)
  \right\}^\omega\psi
  \right]\left(\vec x\right)\,d\vec x\,.
  \endalign
$$
After $*$-transforming
proposition 2.3, Equation (3.1) follows from (3.2).  \qed
\enddemo

\enddocument